\documentclass[10pt,twocolumn,letterpaper]{article}

\usepackage[pagenumbers]{cvpr} 

\usepackage{graphicx}
\usepackage{amsmath}
\usepackage{amssymb}
\usepackage{booktabs}
\usepackage{balance}

\usepackage[pagebackref,breaklinks,colorlinks]{hyperref}

\usepackage[capitalize]{cleveref}
\crefname{section}{Sec.}{Secs.}
\Crefname{section}{Section}{Sections}
\Crefname{table}{Table}{Tables}
\crefname{table}{Tab.}{Tabs.}

\def\cvprPaperID{14} 
\def\confName{CVPR}
\def\confYear{2022}

\begin{document}

\title{PO-ELIC: Perception-Oriented Efficient Learned Image Coding
}

\author{Dailan He\thanks{Equal contribution.}, Ziming Yang\footnotemark[1], Hongjiu Yu\footnotemark[1], Tongda Xu, Jixiang Luo,\\
Yuan Chen, Chenjian Gao, Xinjie Shi, Hongwei Qin\\
SenseTime Research\\
{\tt\small \{hedailan, yangziming, yuhongjiu\}@sensetime.com}
\and
Yan Wang\thanks{Corresponding author.}\\
SenseTime Research\\
Tsinghua University\\
{\tt\small wangyan1@sensetime.com}\\
{\tt\small wangyan@air.tsinghua.edu.cn}
}
\maketitle

\begin{abstract}
   In the past years, learned image compression (LIC) has achieved remarkable performance. The recent LIC methods outperform VVC in both PSNR and MS-SSIM. However, the low bit-rate reconstructions of LIC suffer from artifacts such as blurring, color drifting and texture missing. Moreover, those varied artifacts make image quality metrics correlate badly with human perceptual quality. In this paper, we propose PO-ELIC, \ie, Perception-Oriented Efficient Learned Image Coding. To be specific, we adapt ELIC, one of the state-of-the-art LIC models, with adversarial training techniques. We apply a mixture of losses including hinge-form adversarial loss, Charbonnier loss, and style loss, to finetune the model towards better perceptual quality. Experimental results demonstrate that our method achieves comparable perceptual quality with HiFiC with much lower bitrate.
\end{abstract}

\section{Introduction}
Learned image compression (LIC) has outperformed traditional methods like JPEG~\cite{jpeg} and BPG~\cite{bpg} in terms of PSNR and MS-SSIM. In 2018, the classical hyperprior framework ~\cite{imagecnn5, imagecnn7} dramatically improves the rate distortion performance of LIC. More recently, various context models~\cite{imagecnn6,he2021checkerboard} have been proposed to accurately predict the distribution of latents, so as to further reduce bitrate.
Although these models perform well on full-reference metrics, the reconstructed images show various artifacts when bpp is low (\eg $\leq0.3$). For example, it is well-known that MSE-optimized models produce blurry reconstruction images. The similar phenomenon occurs when optimizing MS-SSIM and other metrics. Those artifacts become increasingly intolerable as bpp grows even lower (\eg $0.075$). In fact, no full-reference metric is fully consistent with perceptual quality, and optimizing towards any of the metrics brings visual artifacts. This is known as perception-distortion trade-off \cite{blau2018perception}.

To address this issue, previous works introduce generative adversarial network (GAN)\cite{goodfellow2014generative} to enhance perceptual quality. \cite{agustsson2019generative} efficiently compress images at low bit-rate and maintain image details by introducing adversarial training. HiFiC~\cite{mentzer2020high} exploits generator and conditional discriminator architectures for perceptual quality. However, to some extent they all face common GAN problems, such as unnatural texture and drifted color. To tackle these challenges, we follow these existing approaches and further investigate the perceptual optimized LIC. Our target is to encode images in lower bitrates with higher perceptual quality.

\begin{figure}
    \centering
    \includegraphics[width=0.99\linewidth]{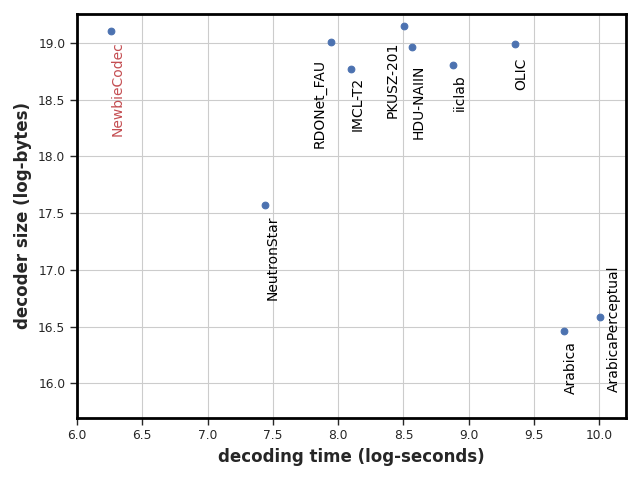}
    \caption{Logarithm decoder size and decoding time of CLIC2022 \textit{Image 075}. Conventional coders (JPEG, BPG and AVIF) are omitted. Our method (red) is the fastest among learning based approaches even with relative large decoder. \textit{Image 150/300} have the similar trend.}
    \label{fig:decoder_time}
\end{figure}

In this paper, we contribute in two aspects:
\begin{itemize}
    \item We propose PO-ELIC, which can utilize lower bit-rate to achieve comparable visual quality against previous approaches. The reconstructions below $0.15$bpp still retain clear and realistic details (See Fig.~\ref{fig:2HiFic} and Fig.~\ref{fig:2HiFic2}).
    \item We exploit the advantage of GAN at low bit-rate and context model at medium bit-rate to balance distortion and rate in Sec.~\ref{architecture}. And Fig.~\ref{fig:decoder_time} shows we have the fastest decoder among other learning based methods on CLIC 2022 leaderboard.
\end{itemize}

\begin{figure*}[tb]
    \centering
    \includegraphics[width=0.85\linewidth]{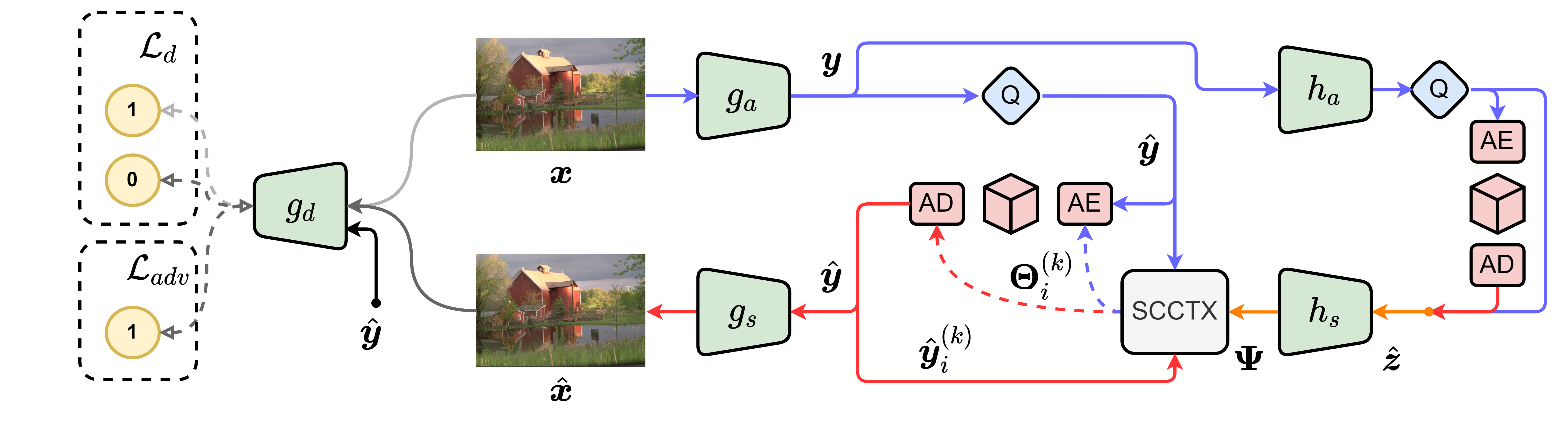}
    \caption{Diagram of the adopted networks. The right part is ELIC~\cite{elic}. We use the same architecture of $g_a, g_s, h_a$ and $h_s$ as the original paper. SCCTX denotes the spatial-channel context model. We use the uneven 5-group scheme with parallel context models~\cite{he2021checkerboard}. The left part shows the adversarial training. We use the same discriminator ($g_d$) structure as HiFiC~\cite{mentzer2020high}.}
    \label{fig:elic-arch}
\end{figure*}

\section{Background}
\subsection{LIC with context model}
Lossy image compression aims to optimize the rate distortion function $\mathcal{R} + \lambda \mathcal{D}$. Denoting the image as $x$, encoder as $g_a$ and decoder as $g_s$, the neural network has the following objective:
\begin{equation}
\mathcal{L}= \mathbb{E}[-\log p(g_a(x)) + \lambda d(x, g_s(g_a(x)))]     
\end{equation}\label{rd_loss}
where $\mathbb{E}$ is the expectation over $p(x)$, $g_a$ extracts the input image $x$ as latent variable $\hat y = g_a(x)$ and $g_s$ transforms it into reconstruction $\hat{x}$. $\mathcal{D}, \mathcal{R}$ are the MSE reconstruction loss and bit-rate computed via learned prior.

Auto-regressive context model is the key factor to promote compression performance by more accurately modeling symbol probability. To be specific, the estimation of current symbol $y_i$ can leverage previous symbols $y_{<i}$:
\begin{equation}
     p(y_i|y_{<i}) = p(y_i| \Psi(y_{<i}))
\end{equation}
where $\Psi$ is context model of various form. Minnen~\etal~\cite{imagecnn7} utilizes spatial masked convolution as context model. Then channel-wise context model is proposed~\cite{minnen2020channel}. ELIC~\cite{elic} adopts a spatial-channel context modelling.

\subsection{LIC with generative adversarial networks}

GAN has been successful in improving perceptual quality of end-to-end image compression~\cite{mentzer2020high, Chen2021MCM, Gao2021clic-huawei}. Usually, a conditional GAN (cGAN) is adopted to constrain the consistency between the decoded image and the original input. The most common adversarial loss of GAN is the non-saturated binary cross-entropy (BCE). Given a discriminator $g_d$, the BCE adversarial loss is:
\begin{equation}
    \mathcal{L}_{adv} = -\mathbb{E}\left[\log g_d(\hat x, \hat y)\right]
    \label{eq:gan-bce-G}
\end{equation}
where the condition $\hat y = g_a(x)$ is the coding-symbols, according to Eq.~\ref{rd_loss}.  By optimizing $g_a, g_s$ guided by $g_d$ we constrain the reconstruction image $\hat{x}$ to be closer to the original one. To train the discriminator $g_d$, an auxiliary discriminator loss is introduced:
\begin{equation}
    \mathcal{L}_{d} = -\mathbb{E}\left[\log g_d(x, \hat y)\right] - \mathbb{E}\left[\log \left(1-g_d(\hat x, \hat y)\right)\right]
    \label{eq:gan-bce-D}
\end{equation}

Introducing the $\mathcal{L}_{adv}$ term to $\mathcal{D}$ extends the rate-distortion optimization to rate-distortion-perception optmization, as GAN demonstrates better correlation with human perception.

\section{Architecture}\label{architecture}

We use ELIC~\cite{elic} as our coding architecture. Fig.~\ref{fig:elic-arch} shows its diagram. When optimizing MSE, it achieves better RD performance than VVC~\cite{bross2021overview} w.r.t. both PSNR and MS-SSIM. The model adopts a multi-dimension context model SCCTX, recognizing redundancy in latents from both channel and spatial dimensions. Because of the usage of parallel context model~\cite{he2021checkerboard}, it gets rid of slow serial decoding and can decompress a 720P image within $100$ms.

\section{Objective}

We take the rate-constrained RD optimization from HiFiC:
\begin{equation}
    \mathcal{L} = \mathcal{D} + \lambda(\mathcal{R}, \mathcal{R}^*) \cdot \mathcal{R}
\end{equation}
where $\mathcal{D}$ and $\mathcal{R}$ are (perceptual) distortion and rate terms. The multiplexer $\lambda(\mathcal{R}, \mathcal{R}^*)$ is conditioned on the given target bitrate $\mathcal{R}^*$:
\begin{equation}
    \lambda(\mathcal{R}, \mathcal{R}^*) = \left\{ \begin{aligned} 
        \lambda_\alpha,\quad \mathcal{R} \ge \mathcal{R}^* \\
        \lambda_\beta ,\quad \mathcal{R} < \mathcal{R}^*
    \end{aligned}\right.
\end{equation}

Our summarized perceptual $\mathcal{D}$ loss function is:
\begin{equation}
\mathcal{D} = \lambda_1 \mathcal{L}_{perc} + \lambda_2 \mathcal{L}_{recon} + \lambda_3  \mathcal{L}_{adv} + \lambda_4 \mathcal{L}_{sty}
\label{eq:total-loss}
\end{equation}
where the perceptual loss $\mathcal{L}_{perc}$ is LPIPS-VGG~\cite{zhang2018unreasonable}.
$\mathcal{L}_{recon}$ is a pixel-wise reconstruction loss ($L_2$, $L_1$, Charbonnier loss~\cite{Lai2017chabonnier}, \etc). $\mathcal{L}_{adv}$ is the adversarial loss, and $\mathcal{L}_{sty}$ is the style loss constraining the texture consistency. Similar loss functions have been successfully used in low-level tasks like image translation~\cite{brock2018biggan} and super-resolution~\cite{sajjadi2017enhancenet}.
We will discuss these loss terms in detail in this section. 

\begin{table*}[ht]
\centering
\caption{Objective results at $0.075, 0.15$ and $0.30$bpp with validation dataset.  ↑ means higher is better and ↓ vice versa.}
\scalebox{1}{
\begin{tabular}{c c c c c c c c c}
\toprule
BPP   & PSNR↑ & MSSSIM↑ & LPIPS↓ & FID↓ & KID↓ & PieAPP↓ & DISTS↓ & IQT↑ \\ \midrule
0.075 & 27.5324        & 0.9179           & 0.1982          & 33.8917       & -0.0286       & 0.7560           & 0.0480          & 0.6783        \\
0.15  & 30.2501        & 0.9424           & 0.1604          & 23.5175       & -0.0292       & 0.4905           & 0.0325          & 0.7136        \\ 
0.3   & 32.6412        & 0.9720           & 0.1083          & 13.9438       & -0.0298       & 0.3788           & 0.0207          & 0.7377        \\ \bottomrule
\end{tabular}\label{subjective_result}
}
\end{table*}

\subsection{Perceptual optimization with SNGAN}

The BCE adversarial loss (eq.~\ref{eq:gan-bce-G} and eq.~\ref{eq:gan-bce-D}) function suffers from the modal collapse issue~\cite{arjovsky2017wgan}. Inspired by ~\cite{miyato2018sngan} and \cite{brock2018biggan}, we instead apply the hinge loss to train a synthesizer with a spectral normalization constrained discriminator:
\begin{equation}
\begin{aligned}
    \mathcal{L}_{adv} =& -\mathbb{E}[g_d(\hat x, \hat y)] \\ 
    \mathcal{L}_d =& -\mathbb{E}[\mathrm{ReLU}(-1+g_d(x, \hat y))]  \\ 
    &- \mathbb{E}[\mathrm{ReLU}(-1-g_d(\hat x, \hat y))]
\end{aligned}
\end{equation}
note that when this hinge loss is used, the output of $g_d$ is the non-activated logits. In our experiments, it outperforms the non-saturated BCE loss.

Other alternatives of BCE adversarial loss include least-square form ~\cite{mao2017LSGAN} and relativistic form ~\cite{jolicoeur2018relativisticGAN}, which are also adopted by recent perceptual LIC approaches~\cite{Gao2021clic-huawei, iwai2021fidelity-controllable}.

\subsection{Learning smoother pixel-wise reconstruction using Charbonnier loss}

$L_1$ loss is frequently used in low-level vision tasks to provide a gentler pixel-wise supervision than $L_2$ (MSE) loss. However, it has an ill-defined gradient when the input is zero. We instead apply a smoother variant of $L_1$ loss called Charbonnier loss~\cite{Lai2017chabonnier}:
\begin{equation}
    \mathcal{L}_{recon}^{(\mathrm{Charb})}(x) = \sqrt{x^2+\epsilon^2},
\end{equation}
where we set $\epsilon=10^{-6}$.

\subsection{Improving texture generation with patched style loss}

Borrowed from style-transfer~\cite{gatys2016style-transfer}, the style loss is widely adopted in low-level tasks to match the texture pattern (or, the so-called \textit{style}) of source and generated images:
\begin{equation}
    \mathcal{L}_{sty}(x, \hat x) = \sum_\ell\left\| G\left(\Phi^{(\ell)}(x)\right) - G\left(\Phi^{(\ell)}(\hat x)\right) \right\| 
\end{equation}
where the operator $G(\cdot)$ denotes the Gram matrix of the given vector. $\Phi$ is the pretrained feature extraction network (\eg, VGG) and $\Phi^{(\ell)}(x)$ is the feature map output by its $\ell$-th selected layer when feed $x$ to the network. The loss matches the global statistics of each feature map, yet the texture usually has locality. As \cite{sajjadi2017enhancenet}, we split the feature maps to $16\times 16$ patches and calculate this loss per patch.

This loss is connected to the LPIPS perceptual loss. In fact, an $1\times 1$ patch style loss is the same as LPIPS without finetuning stage. The LPIPS pays more attention to constraining the global image content and style loss supervises the local texture statistics.

\section{Experiments}

\subsection{Training settings}

We use the ELIC models optimized for MSE as our pretrained models. Following previous works, we use a 8000-image ImageNet subset as training set. To optimize for the objective losses, we train each model for 500 epochs with a batch size of 128. We use Adam optimizer and cosine annealing learning rate scheduler with a base learning rate set to 8e-4.

We finetune the pretrained ELIC model with the above mentioned objective (\ie perceptual loss, reconstruction loss, adversarial loss, and style loss, as summarized in eq.~\ref{eq:total-loss}) to finally obtain the perception-oriented model.

\subsection{Quantitative results}\label{quan_index}
To verify the effectiveness of our method, we utilize \textit{LPIPS}\cite{zhang2018unreasonable}, \textit{FID}\cite{heusel2017gans}, \textit{KID}\cite{binkowski2018demystifying}, \textit{PieAPP}\cite{prashnani2018pieapp}, \textit{DISTS}\cite{ding2020image} and \textit{IQT}\cite{cheon2021perceptual} to guide the evalution of  reconstructions. The combination of these scores is consistent with MOS to some degree. And our major scores are shown in Tab.~\ref{subjective_result}. 

\subsection{Qualitative results}
\begin{figure*}[tb]
    \centering
    \includegraphics[width=.8\textwidth, height=.35\textheight]{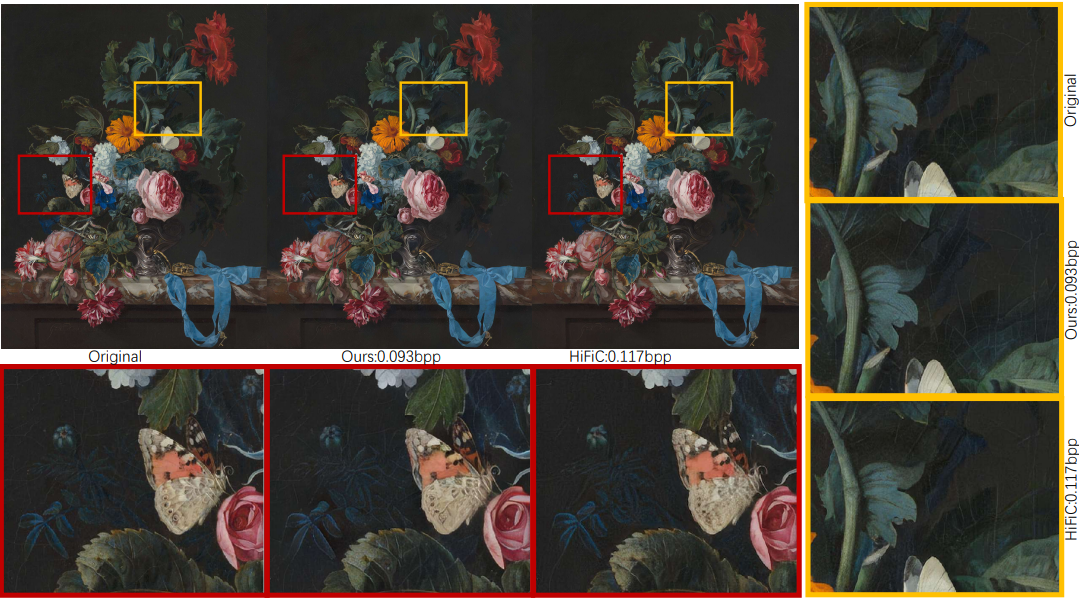}
    \caption{The visualization of our method and HiFiC at low bit-rate. Our method has more details for dark area at right column with yellow rectangles, and more structures on the butterfly at bottom row with red rectangles. }
    \label{fig:2HiFic}
\end{figure*}

\begin{figure*}[h]
    \centering
    \includegraphics[width=.75\textwidth, height=.4\textheight]{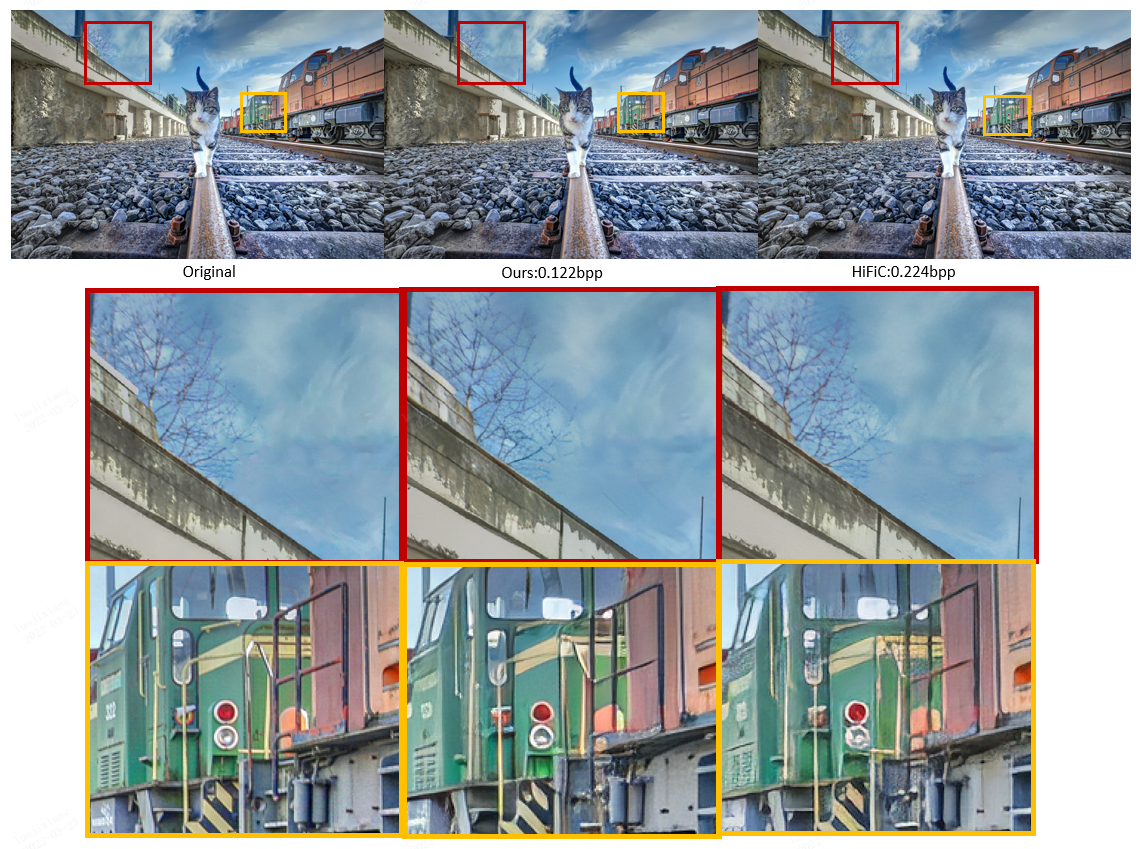}
    \caption{The visualization of our method and HiFiC at low bit-rate. Our method has more details for cable at medium row with red rectangles, and more structures on the train at bottom row with yellow rectangles. }
    \label{fig:2HiFic2}
\end{figure*}

We compare PO-ELIC with HiFiC, and experiments demonstrate that our method has higher fidelity at even lower bit-rate. Fig.~\ref{fig:2HiFic} shows our method has more details for dark area at right column with yellow rectangles, and more structures on the butterfly at bottom row with red rectangles.  Fig.~\ref{fig:2HiFic2} gives another example.

\section{Conclusion}
In this paper we propose PO-ELIC, which introduces the hybrid context and generative model. It utilizes less bits and achieves more pleasant reconstructions compared to HiFiC. Moreover, it further improves the visual quality for LIC at even lower bit-rate ($0.075$bpp). Perceptual metrics such as \textit{LPIPS} and \textit{IQT} indicate that PO-ELIC obtains high-fidelity images with more texture. 

\balance
{\small
\bibliographystyle{ieee_fullname}
\bibliography{ms}
}

\end{document}